\documentclass[USenglish]{lipics-v2021}
\nolinenumbers

\usepackage{color}
\usepackage{xcolor}
\usepackage{hyperref}

\usepackage{listings}
\definecolor{KWColor}{rgb}{0.37,0.08,0.25}
\definecolor{CommentColor}{rgb}{0.12,0.38,0.18}
\definecolor{StringColor}{rgb}{0.06,0.10,0.98}
\definecolor{darkred}{rgb}{0.75,0,0}
\definecolor{lightgrey}{rgb}{0.8,0.8,0.8}

\lstdefinestyle{Eclipse}{
  xleftmargin=0pt,
  basicstyle=\ttfamily\footnotesize,
  commentstyle=\color{CommentColor}\ttfamily\footnotesize,
  stringstyle=\color{StringColor},
  keywordstyle=\color{KWColor}\bfseries,
  escapeinside={/*@}{@*/}
}

\usepackage{algorithm}
\usepackage{algpseudocode}

\usepackage{tabularx}

\usepackage{graphicx}
\graphicspath{ {./images/} }

\usepackage{booktabs}
\usepackage{multirow}
\usepackage{siunitx}

\usepackage{xspace}

\usepackage{wrapfig}

\usepackage{caption}
\usepackage{subcaption}

\usepackage[utf8]{inputenc}
\usepackage[T1]{fontenc}
\usepackage{microtype}

\usepackage{makecell}
\usepackage[numbers,sort&compress]{natbib}

\newif\ifextended
\extendedtrue

\usepackage{tcolorbox}

\definecolor{dkgreen}{rgb}{0,0.3,0}
\definecolor{gray}{rgb}{0.5,0.5,0.5}
\definecolor{dkred}{rgb}{0.5, 0.05, 0.05}
\definecolor{mauve}{rgb}{0.58,0,0.82}
\definecolor{dkgray}{rgb}{0.2, 0.2, 0.2}

\newcommand{\code}[1]{{\textcolor{dkgray}{\texttt{#1}}}}

\newcommand{\toolname}{\textsc{TaintTyper}\xspace}

\newcommand{\taintannot}{\code{@Tainted}\xspace}
\newcommand{\untaintannot}{\code{@Untainted}\xspace}
\newcommand{\polytaint}{\code{@PolyTaint}\xspace}

\newcommand{\codeqlspeeduprange}{2.93X--22.9X\xspace}
\newcommand{\alloptinfpeeduprange}{2.75X--5.52X\xspace}
\newcommand{\locoptinfpeeduprange}{1.47X--2.78X\xspace}
\newif\ifanonymous
\anonymousfalse

\newcolumntype{C}{>{\centering\arraybackslash}X}

\usepackage[capitalise,noabbrev]{cleveref}
\crefname{algocf}{Algorithm}{Algorithms}
\Crefname{algocf}{Algorithm}{Algorithms}

\crefname{algorithm}{Alg.}{Algs.}
\Crefname{algorithm}{Alg.}{Algs.}
\Crefname{section}{Sec.}{Secs.}
\crefname{figure}{Fig.}{Figs.}
\Crefname{figure}{Fig.}{Figs.}
\crefname{line}{line}{lines}
\Crefname{line}{Line}{Lines}

\title{Practical Type-Based Taint Checking and Inference\ifextended{ (Extended Version)}\fi}

\author{Nima Karimipour}{University of California, Riverside, USA}{nima.karimipour@email.ucr.edu}{https://orcid.org/0000-0002-2599-7770}{}

\author{Kanak Das}{University of California, Riverside, USA}{kanak.das@email.ucr.edu}{https://orcid.org/0009-0007-9538-9352}{}

\author{Manu Sridharan}{University of California, Riverside, USA}{manu@cs.ucr.edu}{https://orcid.org/0000-0001-7993-302X}{}

\author{Behnaz Hassanshahi}{Oracle Labs, Brisbane, Australia}{behnaz.hassanshahi@oracle.com}{https://orcid.org/0009-0006-6639-3056}{}

\authorrunning{N. Karimipour, K. Das, M. Sridharan, and B. Hassanshahi}

\Copyright{Nima Karimipour, Kanak Das, Manu Sridharan, and Behnaz Hassanshahi}

\keywords{Static analysis, Taint Analysis, Pluggable type systems, Security, Inference}

\funding{This research was supported in part by the National Science Foundation under grants CCF-2007024, CCF-2223826, and CCF-2312263, and a gift from Oracle Labs.}

\ccsdesc[500]{Software and its engineering~Software verification and validation}
\ccsdesc[500]{Security and privacy~Software security engineering}

\EventEditors{Jonathan Aldrich and Alexandra Silva}
\EventNoEds{2}
\EventLongTitle{39th European Conference on Object-Oriented Programming (ECOOP 2025)}
\EventShortTitle{ECOOP 2025}
\EventAcronym{ECOOP}
\EventYear{2025}
\EventDate{June 30--July 2, 2025}
\EventLocation{Bergen, Norway}
\EventLogo{}
\SeriesVolume{333}
\ArticleNo{25}

\begin{document}

\maketitle

\begin{abstract}

Many important security properties can be formulated in terms of flows of tainted data, and improved taint analysis tools to prevent such flows are of critical need.  Most existing taint analyses use whole-program static analysis, leading to scalability challenges.  Type-based checking is a promising alternative, as it enables modular and incremental checking for fast performance.  However, type-based approaches have not been widely adopted in practice, due to challenges with false positives and annotating existing codebases.  In this paper, we present a new approach to type-based checking of taint properties that addresses these challenges, based on two key techniques.  First, we present a new type-based tainting checker with significantly reduced false positives, via more practical handling of third-party libraries and other language constructs.  Second, we present a novel technique to automatically infer tainting type qualifiers for existing code.  Our technique supports inference of generic type argument annotations, crucial for tainting properties.  We implemented our techniques in a tool \toolname and evaluated it on real-world benchmarks.  \toolname exceeds the recall of a state-of-the-art whole-program taint analyzer, with comparable precision, and \codeqlspeeduprange faster checking time.
Further, \toolname infers annotations comparable to those written by hand, suitable for insertion into source code.  \toolname is a promising new approach to efficient and practical taint checking.

\end{abstract}

\section{Introduction}

Software security is of critical importance to society, as security vulnerabilities can have severe financial and safety impacts.  Many security vulnerabilities can be described as an undesirable flow of \emph{tainted} data.  For example, program inputs possibly controlled by an attacker are tainted, and should not flow to sensitive program operations without proper sanitization.  \emph{Static taint analysis} aims to automatically discover these dangerous data flows, and many approaches to static taint analysis have been proposed~\cite{Banerjee2023compositional,tripp09taj,arzt14flowdroid,DBLP:conf/issta/HuangDMD15,DBLP:journals/pacmpl/GrechS17,codeqlmain,DBLP:conf/sigsoft/EmmiHJPRSSV21,DBLP:conf/sigsoft/WangWZYG020,DBLP:conf/uss/LivshitsL05,DBLP:conf/icse/WassermannS08,DBLP:conf/fase/HuangDM14}.

We desire a static taint analysis with the following properties:
\begin{itemize}
\item \emph{high precision}, i.e., few false positive reports;
\item \emph{high recall}, i.e., few missed vulnerabilities;
\item \emph{fast running times}, enabling checking during continuous integration (CI) or even on local builds; and
\item \emph{applicability} to existing code bases.
\end{itemize}
Previous approaches do not satisfy all of these properties.
Most existing tools perform whole-program analysis, using inter-procedural alias and dataflow analysis to discover tainted flows.  Such approaches are applicable to extant code and can have high precision and recall, but they are difficult to scale to large programs, which can have millions of lines of code~\cite{Banerjee2023compositional,tripp09taj}.  Such analyses can take hours to complete, making them insuitable for CI deployment or local developer runs.  Incremental analysis techniques can speed re-analyzing code after small changes~\cite{Banerjee2023compositional,DBLP:conf/nfm/CalcagnoDDGHLOP15,DBLP:conf/sigsoft/Szabo23}, but they may miss issues~\cite{stein24interactive} or still consume excessive resources~\cite{DBLP:conf/sigsoft/Szabo23}.%

An alternative approach is type-based checking using type qualifiers~\cite{DBLP:conf/pldi/FosterFA99}, as embodied in pluggable type checkers~\cite{PapiACPE2008,DietlDEMS2011}.  A type-based approach performs \emph{modular} checking for tainted flows, combining intra-procedural analysis with annotations at method boundaries to capture inter-procedural flows.  The Checker Framework~\cite{PapiACPE2008,DietlDEMS2011} includes a type-based tainting checker~\cite{cf-manual-taint-checker}.  Such approaches can have high recall, and they have fast running times, as the use of annotations \emph{completely obviates} the need for inter-procedural analysis.  They are also naturally incremental due to modularity, yielding an automatic and significant speedup (nearly 10X in our initial measurements) for small code changes with modern build systems~\cite{gradleIncremental}.  However, extant type-based approaches lack both precision and applicability.  For precision, the previous checker~\cite{cf-manual-taint-checker} treats third-party libraries and various language constructs too conservatively, leading to excessive false positives.  For applicability, running a type-based approach on extant code requires adding annotations, a significant up-front effort that limits adoption.

In this paper, we present a new approach to type-based checking for taint properties that achieves all of our desired properties: high precision and recall, fast running times, and applicability.  In contrast to the extant approach~\cite{cf-manual-taint-checker}, we design our checker to prioritize usability and flexibility over full soundness.  We handle un-annotated library methods as \emph{polymorphic} by default~\cite{DBLP:conf/issta/HuangDMD15}, i.e., only returning tainted data when it is passed in, dramatically reducing the need to annotate such methods.  We also introduce more practical handling of a variety of language constructs to reduce false positives.  While these techniques can in theory introduce unsoundness, our extensive evaluation showed that the impact on analysis recall was minimal.

We also present a novel technique to automatically infer tainting type qualifiers, achieving applicability to existing code.  Our technique extends a recent search-based approach for inferring nullability annotations~\cite{DBLP:conf/sigsoft/KarimipourPCS23} with several important new features.  We develop a novel algorithm for inferring type annotations on generic type arguments, crucial for tainting; this problem was left open in recent work~\cite{DBLP:conf/sigsoft/KarimipourPCS23,DBLP:conf/kbse/KelloggDNAE23}.  Our algorithm can also infer polymorphic method annotations when generic types are not present.  We introduce an optimized handling of local variable annotations to further improve inference performance. 

We implemented our technique in a tool \toolname and evaluated it on a range of benchmarks.  \toolname was able to infer annotations comparable to human-written annotations and suitable for insertion into source code.  Further, once inference was completed, \toolname had higher recall than a state-of-the-art taint analyzer on real-world benchmarks, with comparable precision, and \codeqlspeeduprange faster checking time.  Hence, the evaluation showed that \toolname achieved the precision, recall, speed, and applicability goals outlined above.  An ablation experiment showed that our new checker and inference features were crucial for \toolname's effectiveness.

This paper makes the following key contributions:%
\begin{itemize}
  \item We present a new, more practical type-based tainting checker, with improved handling of third-party libraries and other language constructs.
  \item We describe a novel inference technique for tainting types.  Crucially, the technique handles generic type arguments, and includes an important new optimization to reduce inference time.
  \item We present a detailed experimental evaluation showing our approach makes type-based taint checking much more practical for real-world projects, significantly improving over the state-of-the-art.
\end{itemize}
The implementation of \toolname is available at \url{https://github.com/ucr-riple/TaintTyper}.

\section{Background}\label{sec:background}
In this section, we provide background on typical approaches to taint analysis (\Cref{sec:taint-analysis-background}) and the type-based approach (\Cref{sec:type-based-taint-analysis}), comparing them and motivating our techniques.

\subsection{Taint Analysis}\label{sec:taint-analysis-background}

Taint analysis aims to discover \emph{information flow vulnerabilities} in code~\cite{article:Denning-Denning-InformationFlow}, an undesirable flow of information from some designated \emph{source} of values to a designated \emph{sink} operation.  Our attacker model assumes that the attacker controls inputs originating from untrusted sources and seeks to reach sensitive sinks. In this context, sources include typical mechanisms that receive data from external origins, such as reading from the network, accessing files, or handling user input. Conversely, sinks are operations that have security-critical effects, such as writing to a file, sending data over the network, or executing system or database commands. E.g., for an SQL injection attack, the source is an input from a potential attacker and the sink is a database query, while for a cross-site scripting (XSS) attack, the sink renders output to a web page.
Information flow may be \emph{direct}, solely involving data dependencies, or \emph{indirect}, potentially involving control dependence.  As with most practical static taint analyses, we consider only direct information flow in this work.

The most common approach to static taint analysis is to perform an inter-procedural (whole-program) dataflow analysis to discover source-to-sink data flows~\cite{Banerjee2023compositional,tripp09taj,arzt14flowdroid,DBLP:conf/issta/HuangDMD15,DBLP:journals/pacmpl/GrechS17,codeqlmain,DBLP:conf/sigsoft/EmmiHJPRSSV21,DBLP:conf/sigsoft/WangWZYG020,DBLP:conf/uss/LivshitsL05,DBLP:conf/icse/WassermannS08,DBLP:conf/fase/HuangDM14}.  For languages like Java, inter-procedural analysis requires computing a call graph, typically using whole-program pointer analysis~\cite{sridharan13alias}.  Tracking of tainted data flows also requires handling of pointer aliasing, to account for flows through object fields and data structures like lists.  Precise reasoning about call graphs and pointer aliasing is the key scalability bottleneck for whole-program taint analysis.  For example, a recent approach~\cite{DBLP:conf/pldi/HassanshahiRKSL17} required over 3 hours and 153GB of RAM to precisely analyze the full Java standard library, and real-world programs may grow much larger.  Approaches to make such analyses incremental have been proposed~\cite{Banerjee2023compositional,DBLP:conf/nfm/CalcagnoDDGHLOP15,DBLP:conf/sigsoft/Szabo23}, which could speed up re-analysis of a code base after a small change.  But, such approaches can miss issues~\cite{stein24interactive} or still require significant time or memory~\cite{DBLP:conf/sigsoft/Szabo23}.

Tainting rules capturing which source-sink flows may be vulnerable are typically provided as input to a taint analysis tool.  Such rules may also capture information about \emph{validator} and \emph{sanitizer} operations, whose use may make a source-sink flow safe.  Discovering tainting rules is itself non-trivial and has been a subject of much research~\cite{10.14722/ndss.2014.23039,10.1145/3314221.3314648,10.1145/3510457.3513048,10.1145/1542476.1542485}.  In this work, we focus on the core analysis problems and assume tainting rules have been provided; any improvements in taint rule discovery could be easily combined with our approach.

\subsection{Type-Based Taint Analysis}\label{sec:type-based-taint-analysis}

\subsubsection{Basics} Type-based taint analysis is built on the ideas of pluggable type systems~\cite{PapiACPE2008,DietlDEMS2011}.  A pluggable type system defines a set of \emph{type qualifiers} that refine the built-in types of a language, further restricting the kind of value an expression may evaluate to.  We write type qualifiers by prefixing with an \code{@} character, using Java annotation syntax.  For tainting, the main qualifiers are \code{@Tainted}, for expressions that \emph{may} be influenced by (data-dependent on) a source, and \code{@Untainted}, for expressions that \emph{must not} be influenced by a source.  Tainting rules are provided in the form of \code{@Tainted} qualifiers on source operations and \code{@Untainted} qualifiers on sink operations.%

A pluggable type system must also define a subtyping relationship $\leq$ between qualifiers.  For tainting, we have $\untaintannot\ T \leq \taintannot\ T$ for any base type $T$; \untaintannot values may safely flow to (possibly) \taintannot locations, but not vice-versa.
Enforcing this property requires checking for subtype compatibility at all pseudo-assignments in the program, including assignments to variables / fields, parameter passing, and returns.  To reduce the number of explicit annotations required, pluggable type systems interpret unqualified types as having a default qualifier.  For tainting, the default qualifier is \code{@Tainted}.
The following example illustrates this checking:

\nolinenumbers
\begin{lstlisting}[
    numbers=left,
    firstnumber=1,
    numberstyle=\scriptsize,
    numbersep=5pt
  ]
class Ex {
  @Tainted String f1; @Untainted String f2;
  void m1(@Untainted String s) {
    f1 = s; // no error: @Untainted assigned to @Tainted
  }
  void m2(@Tainted String t) {
    m1(t); // error: @Tainted passed to @Untainted
    f2 = t; // error: @Tainted assigned to @Untainted
  }
}
\end{lstlisting}

In the code, \code{@Tainted String f1} could have been written as just \code{String f1} due to defaulting.  Local variables are not subject to defaulting: their types are inferred using flow-sensitive dataflow analysis~\cite{PapiACPE2008,DietlDEMS2011}, which also enables handling of taint validators.  By default, the Checker Framework's Tainting Checker~\cite{cf-manual-taint-checker} treats all code, including unannotated third-party libraries, with the same defaulting rules.  So, since \taintannot is the default qualifier, any value returned by a third-party method is assumed to be tainted.  This approach leads to too many false positives in practice; \Cref{sec:unannotated-code-handling} describes our more practical handling of such code.

\subsubsection{Method calls} Pluggable type systems must also check that method overriding respects the subtyping rules for the type qualifiers.  Return types must be covariant in overriding methods, and parameter types must be contravariant.  Consider the following (erroneous) example:

\nolinenumbers
\begin{lstlisting}[
    numbers=left,
    firstnumber=1,
    numberstyle=\scriptsize,
    numbersep=5pt
  ]
class Super {
  @Untainted String foo() { return "hi"; }
}
class Sub extends Super {
  // invalid override!
  @Tainted String foo() { return source(); }
}
\end{lstlisting}

\code{Sub.foo()} cannot be allowed to return a \code{@Tainted String}, since code written to handle \code{Super} objects may expect \code{foo()} to return an \code{@Untainted String}.  Without this checking, the following vulnerable code would pass the checker:

\nolinenumbers
\begin{lstlisting}[
    numbers=left,
    firstnumber=1,
    numberstyle=\scriptsize,
    numbersep=5pt
  ]
void process(Super s) { sink(s.foo()); }
process(new Sub());
\end{lstlisting}

Given this method override checking, method calls can be handled in the pluggable types approach \emph{without computing a call graph}.  At a call $c$ to a method with declared target \code{P.m}, the checker needs only to check that the parameters passed and return value use at $c$ are consistent with the type of \code{P.m} itself.  If at runtime, an overriding method \code{Q.m} is invoked at $c$, override checking ensures that the tainting behavior of \code{Q.m} is consistent with \code{P.m}, so no vulnerability is possible.  This ability to check calls without a call graph yields a huge scalability benefit for the pluggable types approach.

\subsubsection{Data structures / polymorphism}\label{sec:background-polymorphism} Pluggable type systems use parametric polymorphism to support storing differing types of data in different instances of a data structure.  Consider the following example:

\nolinenumbers
\begin{lstlisting}[
    numbers=left,
    firstnumber=1,
    numberstyle=\scriptsize,
    numbersep=5pt
  ]
List<@Tainted String> taintedStrs = new ArrayList<>();
List<@Untainted String> untaintedStrs = new ArrayList<>();
taintedStrs.add(source()); untaintedStrs.add("safe");
sink(taintedStrs.get(0)); // error reported
sink(untaintedStrs.get(0)); // no error reported  
\end{lstlisting}

Here we have two \code{ArrayList} instances, one holding possibly-tainted strings and the other holding untainted strings.  The taintedness of list contents is captured by adding a taint qualifier to the generic type argument, e.g., \code{List<@Tainted String>}.  With these type declarations, the pluggable type checker reports an error at the first \code{sink} call while also proving that the second \code{sink} call is safe.

To achieve similar precision, a whole-program static analysis must use \emph{context sensitivity}: each call to the relevant \code{ArrayList} methods must be analyzed separately, and \code{ArrayList}'s internal state must be represented with a context-sensitive heap abstraction~\cite{sridharan13alias}.  Context sensitivity significantly increases the running time of such analyses.  The pluggable types approach uses qualified type arguments to achieve similar precision with much less cost.

\polytaint annotations can be used for polymorphic methods that do not already use type variables.  Consider, e.g., the \code{parentPath} function in \Cref{fig:motivating-example}.  Here, the return value of a call to \code{parentPath} should only be considered \code{@Tainted} if the parameter for that call is \code{@Tainted}; the \polytaint annotations capture this property.  Again, a standard whole-program analysis can achieve this precision with context sensitivity, but at greater cost to scalability.

It is possible that different objects of a non-generic class vary in terms of whether tainted data is stored in their fields.  Such cases can be handled in whole-program analysis using field sensitivity, which models each field of each (abstract) object separately.  Typically, no such modeling is used in the pluggable types approach, outside of generic types.  We have found that this rarely leads to false positives; qualified generic type arguments and \polytaint nearly always suffice for good precision in practice.

\subsubsection{Benefits} A typechecking approach to tainting scales well since the checking is fully modular: each method can be checked in isolation given only the type signatures of fields it accesses and methods it invokes.  Checking is also incremental: after a change, only the code that needs to be re-compiled needs to be re-checked.  Beyond improved scalability, a type-based approach to taint checking has various other advantages~\cite{PapiACPE2008,DietlDEMS2011}.  Qualifiers serve as machine-checked documentation of tainting properties and invariants, and can also aid in safely performing refactorings.  And, errors from the type-based approach can be more understandable, since they only require reasoning about local code and the types of related functions, not an inter-procedural trace. Because our analysis is type-based, it operates over type annotations rather than runtime behavior, meaning the attacker’s knowledge of the program’s internals does not fundamentally affect the analysis. We assume the attacker cannot modify the program’s source code.

Currently, adopting type-based taint checking imposes a significant burden on programmers, as they must manually annotate their own code and any relevant third-party code.  Our checking and inference techniques significantly reduce this burden, making the type-based approach more practical.

\section{Motivating Example and Approach}\label{sec:motivating-and-approach}

\begin{figure}
  \begin{lstlisting}[
      numbers=left,
      firstnumber=1,
      numberstyle=\scriptsize,
      numbersep=5pt
    ]
Map<String, $\color{green!50!black}\texttt{+@Untainted}$ String> paths = ...; /*@ \label{li:paths-decl} @*/
$\color{green!50!black}\texttt{+@PolyTaint}$ String parentPath($\color{green!50!black}\texttt{+@PolyTaint}$ String path) { /*@ \label{li:parentPath-decl} @*/
  return path.substring(0, path.lastIndexOf("/")); /*@ \label{li:parentPath-body} @*/
}
void exec(String name) {
  String path = paths.get(name); /*@ \label{li:paths-get} @*/
  String parent = parentPath(path); /*@ \label{li:parentPath-call} @*/
  $\color{red!50!black}\texttt{sink}$(Paths.get(parent).toAbsolutePath().toFile()); /*@ \label{li:sink-call} @*/
}
void sink(@Untainted String t) { ... }  /*@ \label{li:sink-method} @*/
\end{lstlisting}
\caption{Motivating example for inference.  Green text indicates where annotations are inserted by \toolname.}
\label{fig:motivating-example}
\end{figure}

\subparagraph{Example} \Cref{fig:motivating-example} gives a motivating example for our approach.  (Disregard the green inferred annotations for the moment.)  The \code{sink} method (line \ref{li:sink-method}) requires an \untaintannot argument.  Assuming the values in the \code{paths} map cannot be tainted, the \code{sink} call at line \ref{li:sink-call} is safe.  We shall show how the enhanced checker and inference of \toolname can automatically annotate and validate this example.

The first challenge is the use of library methods like \code{Paths.get}, \code{toAbsolutePath}, and \code{toFile} on line \ref{li:sink-call}.  Without manual annotation, the previous type-based approach~\cite{cf-manual-taint-checker} treats these calls as returning a \taintannot value by default.  Our approach treats most unannotated methods as polymorphic~\cite{DBLP:conf/issta/HuangDMD15}: it assumes they only return \taintannot data when a parameter is \taintannot.  Therefore, by default, \toolname determines that the expression passed to \code{sink} can only be \taintannot if the \code{parent} variable is \taintannot, a correct handling for this example.

The \code{parent} variable is assigned the result of calling \code{parentPath(path)} (line \ref{li:parentPath-call}), where \code{path} is a value in the \code{paths} map (line \ref{li:paths-get}).  As noted above, the values in the \code{paths} map are not tainted, which is captured by giving \code{paths} the type \code{Map<String,@Untainted String>} (line \ref{li:paths-decl}); this type makes the \code{path} variable \untaintannot at line \ref{li:paths-get}.  The \code{parentPath} function can only return \taintannot data if its argument is \taintannot, captured with \polytaint annotations (line \ref{li:parentPath-decl}).  With these annotations, the \toolname checker reports no warning for this code, as desired.

\begin{wrapfigure}{r}{0.6\textwidth}
  \vspace*{-1cm}
  \centering
  \includegraphics[width=0.6\columnwidth]{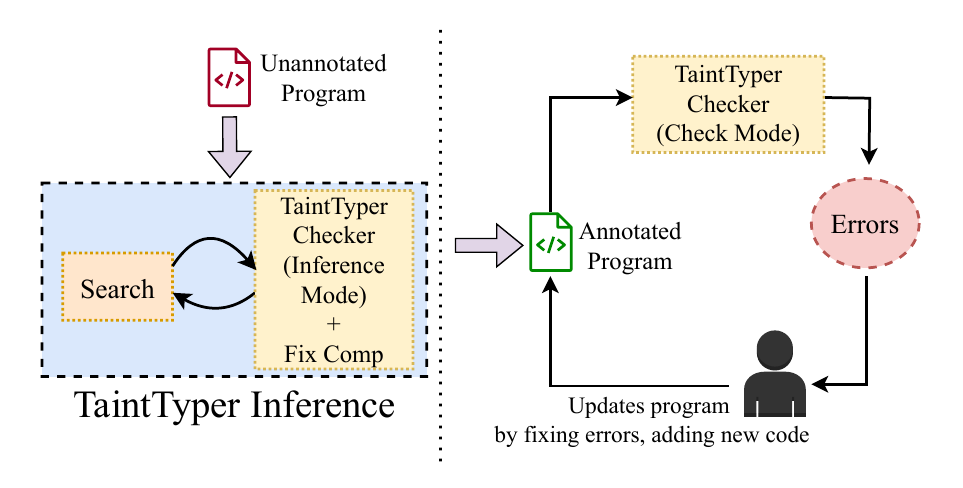}
  \caption{High-level architecture of \toolname.}
  \label{fig:architecture-diagram}
\end{wrapfigure}

The example shows that to be effective, \toolname must support inferring annotations on generic type arguments and \linebreak \polytaint annotations.  The generic type argument support is required to eliminate the error, and \polytaint properly captures the behavior of the \code{parentPath} method.\footnote{For this code excerpt, the parameter and return of \texttt{parentPath} could be marked \texttt{@Untainted}, but this would unnecessarily force all other callers of \texttt{parentPath} to pass in \texttt{@Untainted} data.}  The inference technique of \Cref{sec:inference} can successfully infer the necessary green annotations in \Cref{fig:motivating-example}.

\subparagraph{Our approach} \Cref{fig:architecture-diagram} shows the high-level architecture of \toolname.  Given an unannotated program, \toolname first performs inference to create an annotated version of the program.  The inference improves on a recent search-based technique~\cite{DBLP:conf/sigsoft/KarimipourPCS23} that repeatedly runs a checker to determine the best set of annotations to insert.  Here, the checker is our new type-based taint checker, combined with a new fix computation algorithm to enable inference of type argument annotations and \polytaint.  The inference step only needs to run once.  Afterward, developers need only run the \toolname checker.  They can fix the errors initially reported by the checker and also write new code (with annotations) that will be verified by the checker.  Since the checker is type based, it runs much faster than a whole-program static analysis, enabling quick turnaround times, less CI resource usage, and even checking on local builds.

\section{Practical Type-Based Checking}
\label{sec:checker-improvements}

This section details the new features in \toolname's type checker that reduce false positives in real-world code, making the checker practical.  We discuss handling of unannotated code, and then new handling of other language constructs.

\subsection{Unannotated Code}\label{sec:unannotated-code-handling}\toolname specially handles interactions with unannotated, unchecked code, typically written by a third-party. Given the prevalence of third-party libraries lacking taint annotations in modern Java programs, such interactions occur commonly.  By default, the Checker Framework's Tainting Checker~\cite{cf-manual-taint-checker} uses \taintannot annotation for all code,
whether from source or from libraries.  So, all unannotated library methods are assumed to have a \taintannot return type, yielding too many false positives to be usable.

\toolname adopts a \emph{polymorphic by default} handling of calls to unannotated methods, extending a technique from previous work~\cite{DBLP:conf/issta/HuangDMD15}.  This approach treats the return type and all parameter types (including the receiver type) of an unannotated method as if they were annotated as \polytaint (see \Cref{sec:background-polymorphism}).  For calls to such methods, the return value will be treated as \taintannot only if at least one of the actual parameters at the call is \taintannot.  Consider this expression from line \ref{li:sink-call} from \Cref{fig:motivating-example}:

\nolinenumbers
\begin{lstlisting}[
    numbers=none,
    firstnumber=1,
    numberstyle=\scriptsize,
    numbersep=5pt
  ]
  Paths.get($\color{blue}\textit{parent}$).toAbsolutePath().toFile();
\end{lstlisting}

The invoked \code{toFile} method is unannotated, so with our polymorphic treatment, its return value will only be treated as \taintannot if the result of the \code{toAbsolutePath} call is \taintannot.  Since \code{toAbsolutePath} is also unannotated, the process recurses, and the taintedness depends on the result of the \code{Paths.get(\allowbreak\color{blue}{\textit{parent}})} call.  Finally, since \code{Paths.get} is also unannotated, \toolname concludes the taintedness of the whole expression depends on whether \code{$\color{blue}\textit{parent}$} is tainted, as desired for this example.

\begin{figure}[t] %
  \centering
  \includegraphics[width=0.8\columnwidth]{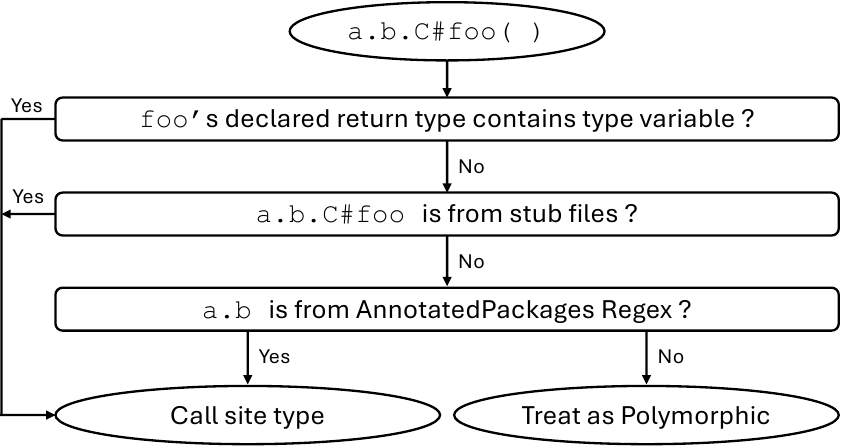}
  \caption{Logic for determining the return type qualifier for a method call, accounting for unannotated code.}
  \label{fig:annotated-unannotated-partition}
\end{figure}

An overview of the logic for determining the return type qualifier for a method call, and when to apply polymorphic defaulting, is given in \Cref{fig:annotated-unannotated-partition} (the logic for parameter types is similar).  The flowchart handles a call to some method \code{a.b.C\#foo}, where \code{a.b} is the package containing class \code{C}.  If \code{foo}'s declared return type contains a generic type variable, we use the standard type checking rule for the call site, even if \code{foo} is in unannotated code.  This exception is important since the return type is at least partially determined by type arguments from the call site.  E.g., line \ref{li:paths-get} of \Cref{fig:motivating-example} invokes \code{Map.get}, whose return type is the type variable \code{V} for map values.  Though \code{Map.get} is unannotated, \toolname applies standard checking, to preserve information from the type argument \code{@Untainted String} given for this particular \code{Map} at line \ref{li:paths-decl}.

\toolname also always uses the standard call site type if \code{foo}'s type been specified in a stub file~\cite{cf-manual-stub-files}, as such files allow for externally providing types for library routines like sources and sinks.  Otherwise, to determine whether code should be treated as annotated, \toolname takes as input an \emph{AnnotatedPackages} regular expression (borrowing from the NullAway nullness checker~\cite{banerjee19nullaway}) which specifies which Java packages should be treated as annotated code.  Note that this setting does not distinguish between first-party and third-party code, providing flexibility.  E.g., when adopting \toolname, source packages can be set as annotated gradually, initially leaving other source packages as unannotated and unchecked.
Only if the \code{a.b} package is not part of the annotated packages, the  polymorphic handling for unannotated code is applied.

The polymorphic-by-default handling of unannotated code is not sound for all cases.  For example, if an unannotated method is itself a tainted source, then its return value should be treated as tainted, independent of the argument taintedness.  Similar issues arise for unannotated sink methods.  As discussed in \Cref{sec:taint-analysis-background}, we view discovery of source and sink methods as a separate problem from the core checking and inference issues we address here.  A method may also return tainted data even with untainted arguments, if a setter method of the receiver is invoked earlier with tainted data.  When discovered, such cases can be addressed via stub file models~\cite{cf-manual-stub-files}.  In practice, our evaluation (\Cref{sec:evaluation}) showed that this technique did not lead to false negatives in comparison to two state-of-the-art tools on real-world benchmarks, and that it dramatically increased practicality compared to~\cite{cf-manual-taint-checker}.

\subsection{Other Constructs}\label{sec:other-language-constructs} \toolname computes a default \untaintannot type for a variety of language constructs that are treated as \taintannot by the previous checker~\cite{cf-manual-taint-checker}.  Enum constants, class literals, lambda expressions, and fields of annotations are always treated as \untaintannot.  A static final field is treated as \untaintannot if its initializer expression is \untaintannot.  An array initializer expression (e.g., \code{new String[] \{x,y\}}) has \untaintannot contents by default if all the initial array values are \untaintannot.  Similarly, a cast expression is \untaintannot if the casted expression is \untaintannot.  In some cases, like static final fields, an explicit \untaintannot annotation could be written, but our defaulting reduces the annotation burden.  Note that other cases like lambda expressions cannot be directly annotated, and can only be handled with a warning suppression without our approach.

\code{java.util.Collection} data structures are widely used in Java programs, making special-case handling useful.  A \code{Collection} is often constructed directly from another \code{Collection} or array, e.g., \code{new ArrayList<>(otherList)}.  In such cases, if the other \code{Collection} holds \untaintannot elements, \toolname always treats the new \code{Collection} as having \untaintannot elements.  The version of~\cite{cf-manual-taint-checker} we compared with did not handle these cases correctly due to type inference limitations.\footnote{The inference limitations have been addressed in more recent versions, but we found that these versions introduced new bugs, so we did not update the version that we compared to.}

The \code{Collection.toArray(T[])} method, which converts a \code{Collection} to a \code{T[]} array, is also frequently used.  With the baseline checker, even for a \code{Collection} of \untaintannot values, the array contents type of a \code{toArray} call is \taintannot unless the \code{toArray} argument is explicitly annotated (e.g., \code{c.toArray(new \untaintannot String[0])}).  \toolname does not require this annotation and treats the result of \code{toArray} as having \untaintannot contents if the \code{Collection} contents are \untaintannot.

\section{Inference}\label{sec:inference}

In this section we detail how \toolname performs inference.  \toolname extends a previous search-based inference technique with support for generic types, \polytaint annotations, and unannotated code.  It also introduces a new optimization that significantly improves inference performance.
 
\subsection{Baseline Algorithm}\label{sec:baseline-algorithm}

Our inference extends the search-based inference approach of Karimipour et al.~\cite{DBLP:conf/sigsoft/KarimipourPCS23}.  The approach aims to infer annotations that minimize the final number of errors reported by a checker, which in~\cite{DBLP:conf/sigsoft/KarimipourPCS23} was the nullness checker NullAway~\cite{banerjee19nullaway}.  By minimizing the final number of errors, this approach infers annotations that \emph{maximize} the amount of code that fully passes the checker.  A recent study showed this search-based approach to work better than alternate approaches in practice~\cite{karimipour25new}.

The search performs \emph{black-box inference}, in which the effectiveness of annotations in reducing errors is measured by running the checker itself and observing its output.  This technique computes a set of annotations that fix some reported checker errors.  Then, the checker is re-applied to see if these annotations cause new errors.  Consider this example:

\nolinenumbers
\begin{lstlisting}[
    numbers=left,
    firstnumber=1,
    numberstyle=\scriptsize,
    numbersep=5pt
  ]
void m1() { m2(source()); m2(source()); }
void m2(String t) { sink(t); }
\end{lstlisting}

Initially, \toolname reports an error at line 2, since \code{t} is \taintannot by default and passed to a sink.  A local fix for this error is to annotate \code{t} as \untaintannot.  But, with this fix, \toolname reports two new errors on line 1 (since \code{source}'s return is \taintannot), increasing the total number of errors.  So, the fix is rejected by the search.  The previous work describes optimizations to speed the search process, by evaluating independent annotations simultaneously~\cite{DBLP:conf/sigsoft/KarimipourPCS23}; we evaluate the effectiveness of these optimizations for taint inference in \Cref{sec:inference-performance}.

The previous work only described how to fix errors reported by NullAway.  For tainting, computing annotations for fixes is significantly more complex due to the need to support generic types and \polytaint.  Recent work on pluggable type inference left open inference for generic types due to the challenges involved~\cite{DBLP:conf/sigsoft/KarimipourPCS23,DBLP:conf/kbse/KelloggDNAE23}.  In \Cref{sec:computing-annotation-fixes}, we describe our novel technique for computing annotations to fix tainting errors, supporting generic type arguments and \polytaint.

\subsection{Computing Fixes}\label{sec:computing-annotation-fixes}

\begin{algorithm}[t]
  \caption{Pseudocode for finding annotations for a fix.}
  \label{alg:infer}
  \begin{small}
    \begin{algorithmic}[1]
  \Procedure{FindAnnots}{$e,T_f$}
  \If{$e$ is a variable $v$}
       \State \Return $ \textsc{UpdateType}(v, T_f) $ \label{li:update-var-type}       
  \ElsIf{$e$ is a binary expression $e_1 \mathit{op}\ e_2$}
       \State \Return $\textsc{FindAnnots}(e_1, T_f) \uplus \textsc{FindAnnots}(e_2, T_f)$ \label{li:binary-op}
  \ElsIf{$e$ is a call $e_1.m(e_2,\ldots,e_n)$} \label{li:start-method-calls}
       \State $G \gets \textsc{GenericsAnnots}(e,T_f)$ \label{li:try-generics}
       \State $\textbf{if}\ G \neq \bot\ \textbf{return}\ G$ \label{li:return-if-generics-succeed}
       \State $P \gets \textsc{PolyTaintAnnots}(e,T_f)$ \label{li:call-polytaint}
       \State $\textbf{if}\ P \neq \bot\ \textbf{return}\ P$ \label{li:return-if-polytaint-succeed}
       \State $m \gets \textsc{InvokedMethod}(e)$  
       \If{$\textsc{TreatAsPolyTainted}(m)$} \label{li:call-treat-poly}
         \State \Return $\biguplus_{i \in \textsc{PTArgs}(m)} \textsc{MakeUntainted}(e_i)$ \label{li:third-party-args-untainted}
       \Else
         \State \Return $\textsc{UpdateType}(\textsc{ReturnType}(m),T_f)$ \label{li:method-call-simple}
       \EndIf \label{li:end-method-calls} 
  \EndIf
  \EndProcedure
  \Procedure{MakeUntainted}{$e$}\label{li:make-untainted-start}
  \State $T' \gets $ \code{@Untainted} $\textsc{TypeOf}(e)$
  \State \Return \textsc{FindAnnots}$(e,T')$
  \EndProcedure\label{li:make-untainted-end}
  \Procedure{GenericsAnnots}{$e,T_f$}
  \State $m \gets \textsc{InvokedMethod}(e)$
  \State $S \gets \textsc{FindTypeSubst}(\textsc{ReturnType}(m),T_f)$ \label{li:find-type-var-subst}
  \State $\textbf{if}\ S = \bot\ \textbf{return}\ \bot$ \label{li:find-type-subst-failed}
  \State $e_r \gets \textsc{ReceiverArg}(e)$ \label{li:get-receiver-arg}
  \State $T' \gets \textsc{ApplySubst}(S,\textsc{ReceiverType}(m),\textsc{TypeOf}(e_r))$ \label{li:apply-subst}
  \State \Return $\textsc{FindAnnots}(e_r, T')$ \label{li:recurse-for-receiver-arg}
  \EndProcedure
  \end{algorithmic}
  \end{small}
\end{algorithm}

For our tainting checker, there are two main causes of reported errors: type incompatibility at a (pseudo-)assignment, and incorrect method overriding (see \Cref{sec:type-based-taint-analysis}).  For both cases, fixing the error requires \emph{adjusting the type} of relevant variables or expressions.  E.g., for the example of \Cref{sec:baseline-algorithm}, the initial fix was to change the type of parameter \code{t} to \code{@Untainted String}.  \Cref{alg:infer} gives our new technique to compute the annotations to achieve a desired type adjustment.  The main procedure \textsc{FindAnnots} takes as arguments an expression $e$ and a desired type for the fix $T_f$.  It either returns a set of annotations that modify $e$'s type to be $T_f$, or $\bot$ to indicate it has failed to find such a set.  \textsc{FindAnnots} relies on various other procedures, some of which we elide for brevity but describe in text.  We also only show handling of key representative language constructs for simplicity. We first explain the basic cases for \Cref{alg:infer}, and then present our novel handling of generic type arguments and \polytaint annotations.

\subsubsection{Basics} For a variable $v$ (which could be a local, parameter, or field), line \ref{li:update-var-type} uses a routine \textsc{UpdateType} (not shown) to update $v$'s declared type directly.  \textsc{UpdateType} may fail and return $\bot$; e.g., \toolname does not attempt to convert a raw type like \code{List} to a generic type like \code{List<@Untainted String>}, as this change is out of scope for our work (other tools can be applied for such cases~\cite{DBLP:journals/toplas/TipFKEBS11}).  For binary operators with sub-expressions $e_1$ and $e_2$, we recursively compute fixes for $e_1$ and $e_2$ and then combine them using a special $\uplus$ operator (line \ref{li:binary-op}).  $\uplus$ propagates the failure value $\bot$---it is defined as follows:
\[
A \uplus B = 
\begin{cases} 
\bot & A = \bot \vee B = \bot, \\
A \cup B & \text{otherwise}.
\end{cases}
\]

Method calls, handled at lines~\ref{li:start-method-calls}--\ref{li:end-method-calls}, require the most complex handling.  We first attempt to handle the call by inferring annotations on generic type arguments (line \ref{li:try-generics}).  If that fails, we attempt inference of \polytaint annotations (line \ref{li:call-polytaint}).  We will explain the generics and \polytaint cases further shortly.  If both generics and \polytaint inference fail (by returning $\bot$), we fall back to a more direct handling, depending on the method being invoked.

line \ref{li:call-treat-poly} checks if the invoked method $m$ should be treated as having \polytaint annotations.  This check returns true if either $m$ has declared \polytaint annotations or if $m$ is unannotated and our default polymorphic handling applies (see \Cref{sec:unannotated-code-handling}).  In such cases, to make the call's type untainted, \emph{all} \polytaint parameters must be made untainted, reflected in line \ref{li:third-party-args-untainted} ($\textsc{PTArgs}(m)$ returns the \polytaint argument positions for $m$).  \textsc{MakeUntainted} (lines~\ref{li:make-untainted-start}--\ref{li:make-untainted-end}) updates the type of expression $e$ with a top-level \code{@Untainted} annotation and then recursively calls \textsc{FindAnnots}.  Any failure to make an argument untainted is propagated using $\uplus$.  Finally, for calls to annotated methods, our base handling is to update the declared return type of the invoked method (line \ref{li:method-call-simple}).

\subsubsection{Generics}\label{sec:generic-fix-computation} For calls to methods involving generic types, our approach aims to find fixes that annotate type arguments instead of directly updating the invoked method's return type.  To see why, consider the call \code{paths.get(name)} on line \ref{li:paths-get} in \Cref{fig:motivating-example}.  This call invokes \code{Map.get}, whose return type is generic type variable \code{V}.  Say that \textsc{FindAnnots} aims to make the result of the call \untaintannot.  The baseline technique for method calls (line \ref{li:method-call-simple} in \Cref{alg:infer}) would change \code{get}'s return type to \code{\untaintannot V}, a valid fix.  But, this change would force \emph{all} calls to \code{Map.get} to return untainted values, preventing \emph{any} \code{Map} from holding possibly-tainted values, which is impractical.

Instead, our technique tries to find a fix that leverages generic type arguments, as shown in the \textsc{GenericsAnnots} routine of \Cref{alg:infer}.  First (line \ref{li:find-type-var-subst}), we call \textsc{FindTypeSubst} (not shown) to find a \emph{substitution} for the type variables in the return type of $m$ that yields the desired type $T_f$.  For the above example, \textsc{GenericsAnnots} is called with $e = \code{paths.get(name)}$ and $T_f = \code{@Untainted String}$. Since \code{get}'s return type is type variable \code{V}, line \ref{li:find-type-var-subst} successfully finds a substitution $S = \code{V} \mapsto \code{@Untainted String}$ that yields $T_f$.  \textsc{FindTypeSubst} may fail to find a substitution, in which case it returns $\bot$ (line \ref{li:find-type-subst-failed}).  \textsc{FindTypeSubst} works by recursing through type structures, mapping type variables to the desired type arguments; we elide details as they are straightforward.

When we successfully find a substitution $S$, we then \emph{apply} $S$ to the declared receiver type of the method, and recursively try to find corresponding annotations for the receiver argument of the call (lines~\ref{li:get-receiver-arg}--\ref{li:recurse-for-receiver-arg}).  If $S$ does not cover all type variables in the receiver type, we reuse the generic type arguments from the receiver argument at the call site.  For our \code{paths.get} example, the declared type of the \code{Map.get} receiver is \code{Map<K,V>}, but our substitution $\code{V} \mapsto \code{@Untainted String}$ does not cover \code{K}.  So, we reuse the \code{String} type argument for \code{K} from line \ref{li:paths-decl} of \Cref{fig:motivating-example}, yielding a recursive call $\textsc{FindAnnots}(\texttt{paths},\texttt{Map<String,@Untainted String>})$ at line \ref{li:recurse-for-receiver-arg}.

The recursive nature of \textsc{FindAnnots} successfully handles much more complex uses of generic types, e.g.:

\nolinenumbers
\begin{lstlisting}[
    numbers=left,
    firstnumber=1,
    numberstyle=\scriptsize,
    numbersep=5pt
  ]
void foo(Map<Integer, $\color{green!50!black}\texttt{+@Untainted}$ String> t) {
  sink(t.values().iterator().next());
}  
\end{lstlisting}

\textsc{FindAnnots} aims to make the return type for the \code{next} call \code{@Untainted String}, but it is not immediately evident which generic type argument must be annotated to achieve this.  In our algorithm, the generics logic proceeds by recursively trying to make the \code{iterator} call return \code{Iterator<@Untainted String>}.  This in turn leads to trying to make the \code{values} call return \code{Collection<@Untainted String>}, which finally leads to successfully adjusting the type of \code{t} to \code{Map<Integer, @Untainted String>}.  \toolname can also annotate generic type arguments in \code{extends} clauses of class declarations, and generic methods (where the type variable is scoped to the method instead of the class) are also handled fully.

\begin{algorithm}[ht]
  \caption{Pseudocode for inferring polymorphic annotations.}
  \label{alg:infer-poly}
  \begin{small}
  \begin{algorithmic}[1]
  \Procedure{PolyTaintAnnots}{$e, T_f$}
  \State $m \gets \textsc{InvokedMethod}(e)$  
  \State $\text{Result} \gets \emptyset$
  \State $F_{\text{return}} \gets \textsc{FindAnnotForReturnStatements}(m, T_f)$
  \State $\text{Worklist} \gets F_{\text{return}}$
  \State $\text{Processed} \gets \emptyset$
  \While{$\text{Worklist} \neq \emptyset$}
      \State $F_{\text{element}} \gets \text{Worklist.pop()}$
      \If{$F_{\text{element}}$ is not on a local variable}
          \State $\text{Result} \gets \text{Result} \cup \{F_{\text{element}}\}$
          \State \textbf{continue}
      \EndIf
      \If{$F_{\text{element}} \in \text{Processed}$}
          \State \textbf{continue}
      \EndIf
      \State $\text{Processed} \gets \text{Processed} \cup \{F_{\text{element}}\}$
      \State $F_{\text{assign}} \gets \textsc{FindAnnotForAssignments}(m, F_{\text{element}}, T_f)$
      \State $\text{Worklist} \gets \text{Worklist} \cup F_{\text{assign}}$
  \EndWhile
  \State $F_{\text{Parameters}} \gets \{ F \mid F \in \text{Result} \wedge F\  \text{is on parameter of}\ m \}$
  \State $F_{\text{NonParameters}} \gets \text{Result} \setminus F_{\text{Parameters}}$
  \If{$F_{\text{Parameters}} = \emptyset$}
      \State \Return $\textsc{MakeUntainted}(m)$
  \Else
      \State $\text{PolyMethodFix} \gets \textsc{MakePolyTainted}(m, F_{\text{Parameters}}) \cup F_{\text{NonParameters}}$
      \State $F_\text{args} \gets \emptyset$
      \For{$\text{arg} \in \text{PolyMethodFix.args}$}
          \State $F_\text{args} \gets F_\text{args} \cup \textsc{FindAnnots}(\text{arg}, \text{UpdatedTarget}(T_f))$
      \EndFor
  \EndIf
  \State \Return $F_\text{args} \biguplus \text{PolyMethodFix}$
  \EndProcedure
  \end{algorithmic}
\end{small}
\end{algorithm}

\subsubsection{\texttt{@PolyTaint} inference} As noted in \Cref{sec:type-based-taint-analysis}, \polytaint can be useful when a method is generic in its tainting behavior but was not declared using generic type variables.  Due to the lack of type variables, inference of \polytaint annotations must \emph{discover} relevant data flow from parameters to return values, which may occur through callee methods.  E.g., for the \code{parentPath} method in \Cref{fig:motivating-example}, the taintedness of the \code{path} argument influences the return taintedness via a call to \code{path.substring}.

The \textsc{PolyTaintAnnots}\label{sec:polytaint-fixes} procedure for inferring \polytaint annotations (called at line \ref{li:call-polytaint} in \Cref{alg:infer}) is conceptually simple: it works by inserting \polytaint annotations, observing where such annotations lead to type checking errors, and then recursively inserting more annotations to fix those errors.  However, we found that a straightforward implementation based on this strategy was too inefficient, so we introduced two improvements.  First, a na\"ive approach to discovering new type errors is to re-run the full type checker, but leads to many expensive checker runs; instead,
we implemented our own limited analysis of data flows relevant to \polytaint to discover new errors.  Second, we bounded the depth of the search into callee methods, giving up and returning $\bot$ if inference required searching further (we found depth five to work best in our experiments).  

\Cref{alg:infer-poly} gives a simplified view of the \textsc{PolyTaintAnnots} procedure. The algorithm starts by invoking \textsc{FindAnnotForReturnStatements} (not shown), which scans the method body for returned expressions and uses \textsc{FindAnnots} to discover the annotations needed to align each expression's type with $T_f$.  If any of the computed annotations targets a local variable $v$, \textsc{PolyTaintAnnots} uses \textsc{FindAnnotForAssignments} (not shown) to scan for expressions assigned to $v$ and compute necessary annotations for those expressions (again via \textsc{FindAnnots}).  This process iterates using a worklist until all locals are handled.
Upon completing the iterations, if there exists an annotation on a method parameter, the method is identified as polymorphic, and the necessary annotations are computed.
The method indirectly invokes \textsc{FindAnnots} from \Cref{alg:infer} through calls to \textsc{FindAnnotForReturnStatements} and \textsc{FindAnnotForAssignments}, which may recursively invoke \textsc{PolyTaintAnnots}.  A separate depth bound ensures termination, and the algorithm returns $\bot$ if the bound is reached or if any call to \textsc{FindAnnots} returns $\bot$.
When a method is determined to be polymorphic, the algorithm calculates the required annotations for its arguments and combines these with the annotations derived for the return type.

\subsubsection{Example}\label{sec:inference-alg-example} We now discuss the overall process of applying inference to our motivating example in \Cref{fig:motivating-example}.  Since the error in the unannotated code is reported at line \ref{li:sink-call}, initially \textsc{FindAnnots} is invoked to try to make the \code{toFile()} call passed to \code{sink} have type \code{@Untainted String}.  As discussed in \Cref{sec:unannotated-code-handling}, this expression includes nested calls to unannotated code, handled by line \ref{li:third-party-args-untainted} in \Cref{alg:infer}.  Eventually, this leads to annotating \code{parent} as \untaintannot.
This annotation causes a new checker error at line \ref{li:parentPath-call}, leading inference to use \textsc{FindAnnots} to make the \code{parentPath(path)} call \untaintannot.  Here, our \polytaint inference succeeds for \code{parentPath}, leading to the annotations on line \ref{li:parentPath-decl}.  The search then makes \code{path} \untaintannot, causing a type error at line \ref{li:paths-get}.  Here, \textsc{FindAnnots} makes the \code{paths.get} call \untaintannot by updating the generic type of the \code{paths} field, discussed in detail in \Cref{sec:generic-fix-computation} above.  With this change, no new errors are reported, completing inference. 

\subsubsection{Local Variable Optimization}\label{sec:local-vars-optimization} Our initial inference implementation took an excessive amount of time, nearly 24 hours for larger benchmarks.  Most inference time is spent running the checker to detect the impact of annotations on warnings.  For tainting, we found that many such checker runs were for annotations on local variables, e.g., the runs for \code{parent} and \code{path} in \Cref{fig:motivating-example} (discussed above).  As an optimization, we enhanced our fix computation to \emph{internally} determine the impact of local variable annotations rather than using the checker.  This reasoning requires finding assignments to the relevant locals, and then recursively invoking \textsc{FindAnnots} to make the type of each assignment's right-hand side match the local's new type, similar to the logic shown in \Cref{alg:infer-poly} for inferring \polytaint.  With this optimization, two checker calls (for \code{parent} and \code{path}) are eliminated in inference for \Cref{fig:motivating-example}.  In \Cref{sec:inference-effectiveness} we show this optimization has a very significant impact on inference performance.

\section{Implementation}\label{sec:implementation}

\toolname includes both type-based taint checking and inference (see \Cref{fig:architecture-diagram}).  \toolname's checker (see \Cref{sec:checker-improvements}) was built using the Checker Framework~\cite{PapiACPE2008,DietlDEMS2011}, version 3.42.0.  Use of the Checker Framework equips the \toolname checker with robust support for flow-sensitive local type inference, checking of generic types, and qualifier polymorphism.  For our prototype, we modeled a number of source and sink methods involved in the most common Java vulnerabilities~\cite{owasp-top-ten}.  Our sinks include common methods that write to a file, send data over the network, or execute sensitive system or database commands.  We modeled sources that read from the network, the file system, or user input.  We also modeled a few relevant well-known sanitizer methods~\cite{findsecbugs-sanitizers}.  As noted in \Cref{sec:background}, creating more complete databases of sources, sinks, and sanitizers is still a research problem, and \toolname can easily benefit from further advances in that area.

\toolname's inference implementation uses a modified version of the \textsc{NullAwayAnnotator} tool~\cite{DBLP:conf/sigsoft/KarimipourPCS23}.  Inside the \toolname checker, we implemented \Cref{alg:infer} to find annotations that can fix an error, and added support for serializing this information in the checker output.  Our inference implementation reads in this serialized output to use during its search.  The search is similar to that of~\cite{DBLP:conf/sigsoft/KarimipourPCS23}, enhanced with our local variable optimization (\Cref{sec:local-vars-optimization}).  The search is depth bounded~\cite{DBLP:conf/sigsoft/KarimipourPCS23}, and we used a bound of 15 in our experiments.

\section{Evaluation}\label{sec:evaluation}

Here we present an experimental evaluation showing the high effectiveness of \toolname in practice.

\subsection{Experimental setup and research questions}\label{sec:experimental-setup}

From previous work we found three benchmark suites will all tainting violations labeled, serving as a ground truth. They are:
\begin{itemize} 
  \item \textbf{Securibench Micro}~\cite{securibenchMicro}, which provides 122 servlets exhibiting potential information-flow vulnerabilities, with the source code annotated with benign or problematic flows.
  \item \textbf{JInfoFlow}~\cite{DBLP:journals/pacmpl/GrechS17}, a self-contained benchmark of 12 information-flow violations featuring reflection-intensive,
  event-driven code without dependencies on external libraries.
  \item \textbf{Injection Experiments}~\cite{DBLP:journals/toplas/SpotoBEFLMS19}, 
  comprising 8 Java programs with information-flow violations reported by their tool. While the original tool is no longer available, the dataset remains accessible. In the metadata, the authors label the reports from their tool as true or false positives. 
\end{itemize}
Although these three benchmarks capture various interesting cases, they consist either of toy examples~\cite{securibenchMicro, DBLP:journals/pacmpl/GrechS17} or projects that have not been maintained for years~\cite{DBLP:journals/toplas/SpotoBEFLMS19}. Consequently, we use them solely as one validation of the impact of the techniques of \Cref{sec:checker-improvements} on \toolname's recall.

For a more realistic evaluation, we examine a suite of actively-developed open-source Java programs used in recent work~\cite{DBLP:conf/pldi/AntoniadisFKRAS20,DBLP:conf/msr/BuiSF22}. We selected only projects for which there were some reported tainting errors.  We also strove to include a variety of project sizes and types, ensuring the benchmark reflects real-world scenarios; our suite includes a web framework, content management system, security framework, forum software, and library management system. Due to the lack of a ground truth on these benchmarks, we made considerable manual efforts to ensure the accuracy of our results. This included carefully comparing results against previous static analyzers and manually annotating the code for comparison (further details below). In the end, we prioritized benchmarks that were the most representative and important, ultimately selecting seven projects for our evaluation, as listed in \Cref{table:oss:info}.  We used all benchmarks from~\cite{DBLP:conf/pldi/AntoniadisFKRAS20} for which there were reported tainting errors except for webgoat and opencms, which were not included due to complex build system configurations that \toolname cannot yet handle. One such complexity arises when the code relies on generated code, such as when using Project Lombok~\cite{lombok}. If a generated getter method is inferred to return untainted, it must first be delomboked and explicitly annotated. However, when the code is compiled, the generated code is overwritten, causing the loss of information needed to propagate the untainted annotation from the getter to the corresponding field. This is not a fundamental limitation of our approach but rather an implementation challenge that could be addressed with additional engineering effort.

\begin{table}
  \footnotesize
  \centering
  {\renewcommand{\arraystretch}{1.2}
  \begin{tabular}{lrrr}
      \toprule
      \multirow{2}{*}{Project} & \multirow{2}{*}{KLoC} & \multicolumn{2}{c}{Inferred Annotations} \\ \cline{3-4} 
                                 &  & Total & per kLoC \\ \hline
      esapi-java-legacy & 18.3 & 265 & 14.4  \\ \hline
      pybbs & 9.7 & 105 & 10.8 \\ \hline
      alfresco-core & 13.4 & 81 & 6.0 \\ \hline
      alfresco-remote-api & 85.8 & 213 & 2.5 \\ \hline
      cxf & 47.5 & 296 & 6.2 \\ \hline
      struts & 49.3 & 313 & 6.3  \\ \hline
      commons-configuration & 20.2 & 261 & 12.9 \\ 
      \bottomrule
  \end{tabular}}
  \caption{Benchmark sizes and inferred annotation counts.}
  \label{table:oss:info}  
\end{table}

We used CodeQL~\cite{codeqlmain} v2.15.1 and P/Taint~\cite{DBLP:journals/pacmpl/GrechS17,doopCode} as baseline tools for our evaluation.  CodeQL is a production-quality security analysis tool, widely used and freely available.  P/Taint uses state-of-the-art whole-program analysis techniques, also employed in recent work~\cite{Banerjee2023compositional}.  We configured both tools to detect issues involving the sources and sinks modeled for \toolname (\Cref{sec:implementation}).
We considered a variety of other tools for our experiments but found them unsuitable.  FlowDroid~\cite{arzt14flowdroid} is a well-known taint analyzer for Android, but it does not support analysis of Java server programs.  And, the taint analyses in SonarQube~\cite{sonarqubeMainPage}, RAPID~\cite{DBLP:conf/sigsoft/EmmiHJPRSSV21}, and \textsc{CompTaint}~\cite{Banerjee2023compositional} are not freely available.

Given these benchmarks and tools, we studied seven research questions:
\begin{description}
   \item[\textbf{RQ1}] \label{item:soundness-existing-benchmarks} Does \toolname find the known errors in existing labeled benchmarks?
   \item[\textbf{RQ2}] \label{item:errors-vs-codeql} After inference, how do \toolname's reported errors compare to those reported by state-of-the-art tools (in terms of precision and recall)?
   \item[\textbf{RQ3}] \label{item:checker-running-time} After inference, how does \toolname's checking time compare with state-of-the-art tools on real-world benchmarks?
   \item[\textbf{RQ4}] \label{item:inference-running-time} Does \toolname's inference run in a reasonable amount of time, and are our optimizations effective?
   \item[\textbf{RQ5}] \label{item:number-of-annotations} Does \toolname require a reasonable number of annotations?
   \item[\textbf{RQ6}] \label{item:human-comparison} How do the annotations inferred by \toolname compare to manually-written annotations?
   \item[\textbf{RQ7}] \label{item:ablation} How is \toolname's effectiveness impacted if we disable checker improvements, generic type argument inference, and \polytaint inference?
\end{description}
We address RQ1--RQ4 in \Cref{sec:inference-effectiveness}.  Then, \Cref{sec:assessing-annotations} addresses RQ5 and RQ6, and \Cref{sec:ablation} answers RQ7.
All experiments were conducted in a Google Cloud instance with an AMD EPYC Milan 3rd Generation 2.45GHz CPU with 32vcpu (16 cores) and 128GB memory.

\subsection{Inference effectiveness}\label{sec:inference-effectiveness}

\begin{table*}[t]
  \scriptsize
  \centering
    \begin{tabular}{lr|lr|lr}
    \hline
    \multicolumn{2}{c}{\textbf{SecuriBench Micro}} & \multicolumn{2}{c}{\textbf{JInfoFlow-bench}} & \multicolumn{2}{c}{\textbf{Injection Experiments}} \\
    \hline
    Total & 136(137) & Total & 12 & Total & 730(745) \\
    aliasing & 12 & JInfoFlow/basic & 2 & Snake\&Ladder & 40 \\
    arrays & 9 & JInfoFlow/ctx & 5 & MediaPlayer& 0 \\
    basic & 61 & JInfoFlow/event & 5 & EmergencySNRest & 0(3) \\
    collections & 14 &  &  & FarmTycoon & 5 \\
    datastructures & 5 &  &  & Abagail & 0 \\
    factories & 3 &  &  & JExcelAPI & 3(4) \\
    inter & 16 &  &  & Colossus & 682(693) \\
    pred & 5 &  &  &  &  \\
    reflection & 4 &  &  &  &  \\
    sanitizers & 3(4) &  &  &  &  \\
    session & 3 &  &  &  &  \\
    strong updates & 1 &  & &  &  \\
    \hline
    \end{tabular}
    \caption{Without inference, \toolname reports all the labeled true-positive issues across these three benchmarks. With inferred annotatations, it misses one issue in Securibench Micro, detects all issues in JInfoFlow, and misses 15 issues in Injection Experiments.}
    \label{tbl:external-validation}
\end{table*}
\subsubsection{Soundness on labeled benchmarks}\label{sec:inference-soundness}

To evaluate the soundness of \toolname, the three labeled benchmark suites were used.  For these experiments, we customized the source and sink specifications used by \toolname to match what was expected by each benchmark suite.%

We first ran \toolname's checker without inference, and confirmed that it did not miss \emph{any} labeled vulnerabilities in the benchmarks. We then applied our inference to annotate the benchmarks automatically and re-check them again, to check how the inferred annotations impacted recall.

For Securibench Micro, the metadata indicated 136 labeled vulnerabilities. However, manual inspection revealed 137 in total, with one unlabeled issue in \textit{Basic26.java} and one missing label in \textit{Basic31.java}.  Securibench Micro also makes extensive use of raw types, which appear rarely in modern Java code and are not yet fully handled by \toolname; we inserted (unannotated) generic type arguments for these cases.  With these fixes, after inference, \toolname only missed one true issue out of 137; this was due to an interaction between our polymorphic-by-default library handling and side effects, as discussed in \Cref{sec:unannotated-code-handling}.

In inference-annotated JInfoFlow-bench, \toolname successfully identifies all 12 labeled vulnerabilities. The Injection Experiments consist of eight Java programs, one of which could not be compiled. We annotated the remaining seven programs (collectively comprising 754 labeled issues) using \toolname inference. Eight of the labeled issues occur in files not part of the available benchmark codebase, and one occurs in test code.  Upon manually 
inspecting the inferred annotations and the reported errors from \toolname, we found that it discovered 730 of the remaining 745 issues, with the missed issues again due to side-effecting third-party library calls.  The number of missed issues across these benchmarks is small, and as shall be shown in \Cref{sec:results-reported-errors}, we saw \emph{no} missed issues for \toolname when compared to a production-quality tool on real-world benchmarks.  \Cref{tbl:external-validation} summarizes our findings across these benchmarks.%

\begin{tcolorbox}[width=\columnwidth, arc=3mm, boxsep=0.25mm]
  \textbf{RQ1:} \toolname's checker identifies all known issues in the labeled benchmarks, and inference reduces recall only slightly.
\end{tcolorbox}

\begin{table*}[t]
  \scriptsize
  \centering
  \renewcommand{\arraystretch}{1.3}
  \begin{tabular}{l|rrrrr}
    \toprule
    \multirow{2}{*}{Project} & \multirow{2}{*}{TP} & \multicolumn{2}{c}{CodeQL} & \multicolumn{2}{c}{\toolname} \\ \cline{3-6}
        &     & Precision & Recall & Precision & Recall  \\ \hline
    esapi-java-legacy  & 18  & 0.90      & 0.50   & 0.95  & 1.00  \\ \hline
    pybbs  & 9   & 1.00      & 0.89   & 1.00      & 1.00  \\ \hline
    alfresco-core  & 2   & 1.00      & 0.50   & 1.00      & 1.00 \\ \hline
    alfresco-remote-api & 21  & 0.82      & 0.43   & 0.70      & 1.00 \\ \hline
    cxf   & 9   & 1.00      & 0.11   & 0.75      & 1.00 \\ \hline
    struts   & 23  & 0.39      & 0.39   & 0.50      & 1.00 \\ \hline
    commons-configuration  & 11  & 1.00      & 0.73   & 0.69      & 1.00 \\ 
    \bottomrule
  \end{tabular}
  \caption{Precision and recall results across our benchmark suite.}
  \label{tbl:precision-recall-runtime}
\end{table*}

\begin{table*}[t]
  \scriptsize
  \centering
  \renewcommand{\arraystretch}{1.3}
  \resizebox{\textwidth}{!}{
    \begin{tabular}{l|rrr|rrrr}
      \toprule
      \multirow{2}{*}{Project} & \multicolumn{3}{c|}{Analysis Runtime} & \multicolumn{4}{c}{Inference Runtime} \\ \cline{2-8}
          &  \toolname & CodeQL  & P/Taint & ALL  & LOC  & CORE  & NONE  \\ \hline
      esapi-java-legacy  & 19s       & 94s     & 53m     & 23m  & 65m  & 47m   & 105m  \\ \hline
      pybbs  & 12s       & 83s     & 35m     & 8m   & 15m  & 11m   & 22m   \\ \hline
      alfresco-core  & 15s       & 75s     & 21m     & 9m   & 14m  & 20m   & 31m   \\ \hline
      alfresco-remote-api &  69s       & 202s    & 167m    & 134m & 311m & 391m  & 711m  \\ \hline
      cxf   & 47s       & 1076s   & 151m    & 148m & 294m & 498m  & 817m  \\ \hline
      struts   & 39s       & 425s    & >48h    & 343m & 629m & 1354m & >48h  \\ \hline
      commons-configuration  & 25s       & 114s    & 209m    & 105m & 168m & 299m  & 432m  \\ 
      \bottomrule
    \end{tabular}
  }
  \caption{Analysis runtime and inference runtime results across our benchmark suite.}
  \label{tbl:runtime}
\end{table*}

\subsubsection{Reported errors}\label{sec:results-reported-errors} \Cref{tbl:precision-recall-runtime} compares the precision and recall of \toolname and CodeQL for our benchmarks, addressing RQ2.  For \toolname, the errors are computed after inference has run, so the code includes inferred annotations.  Computing recall requires knowing the ground truth of all real issues in these benchmarks, which is infeasible to collect.  To estimate recall, we use the union of all true positive issues reported by \toolname, CodeQL, and P/Taint as our ground truth.  (We treat a report as a true positive if the corresponding dataflow is deemed feasible by manual inspection.)  This may over-estimate the true recall of all tools, but it provides a good basis for comparing the tools.  The ``TP'' column gives the number of true positive issues for each benchmark.

Comparing error reports between the tools was non-trivial due to their different reporting approaches.  For \toolname, an error is typically reported as a code location where a \taintannot value is being written into an \untaintannot location.  CodeQL reports an error as a data flow trace from a source to a sink, and P/Taint reports an error as a source/sink pair without a trace.  Comparing these errors required manually matching each true-positive \toolname error to a step in some CodeQL trace, or to some data flow for a P/Taint source/sink pair, and vice versa.  CodeQL sometimes treats a formal parameter as a tainted source, without any explicit call passing in tainted data.  In these cases, \toolname annotates the parameter as \untaintannot, capturing the fact that tainted data should not be passed in, but does not report an error.  We count such cases as equivalent to reporting the error; \toolname could easily report errors for such cases if desired.

We exclude the precision and recall of P/Taint from \Cref{tbl:precision-recall-runtime} as both were very low on our benchmarks.  P/Taint found a total of six true positive issues across the benchmarks (all of which were also found by some other tool),  and it reported 50 false positives.  We carefully checked our P/Taint configuration and confirmed it found expected issues in smaller benchmarks like Securibench Micro~\cite{securibenchMicro}.  We also consulted with the tool authors, who acknowledged that P/Taint may not handle these benchmarks well (e.g., due to missing framework support).  Still, triaging P/Taint results was very useful, to gain further confidence in our ground truth.

\toolname finds all true positive issues discovered by CodeQL and P/Taint, leading to a recall of 1 on all benchmarks in \Cref{tbl:precision-recall-runtime}.  \toolname also finds additional true positives missed by CodeQL, reflected in CodeQL's lower recall numbers.  \toolname has lower precision than CodeQL on three benchmarks; we suspect this is due to heuristics in CodeQL that are missing in \toolname.  Still, \toolname matches or exceeds CodeQL's precision on the other four benchmarks.

\begin{tcolorbox}[width=\columnwidth, arc=3mm, boxsep=0.25mm]
\textbf{RQ2:} On our benchmarks, \toolname has outstanding recall, with comparable precision to CodeQL.
\end{tcolorbox}

\subsubsection{Performance}\label{sec:inference-performance}\Cref{tbl:runtime} gives analysis runtimes for \toolname, CodeQL, and P/Taint checking.  The speedups of \toolname's checker over CodeQL are quite significant, ranging from \codeqlspeeduprange.  And, we see orders-of-magnitude speedups compared to P/Taint, which could not analyze the struts benchmark within a 48-hour time limit.  Given that \toolname's checking is modular and incremental, we expect even larger speedups over the whole-program analysis approach for larger benchmarks.  As a sanity check of the benefits of incremental checking, we manually re-ran \toolname checking for five randomly-chosen source files for alfresco-remote-api and struts (the two largest benchmarks), and observed further speedups of 8.7X and 10.9X respectively.

\begin{tcolorbox}[width=\columnwidth, arc=3mm, boxsep=0.25mm]
\textbf{RQ3:} \toolname's checker runs much faster than the baseline tools, with further incremental speedups.
\end{tcolorbox}
\Cref{tbl:runtime} shows inference performance with all optimizations enabled in the ALL column.  Fully-optimized inference always ran in less than 6 hours, suitable for an overnight run and acceptable since it only needs to run once (see \Cref{fig:architecture-diagram}).  The LOC, CORE, and NONE columns respectively show running times with our new local variable optimization only (\Cref{sec:local-vars-optimization}), the core optimizations of~\cite{DBLP:conf/sigsoft/KarimipourPCS23} only (see \Cref{sec:baseline-algorithm}), and no optimizations.  The core optimizations of~\cite{DBLP:conf/sigsoft/KarimipourPCS23} seem to have a lesser impact for tainting inference than they did for nullability; excluding struts (which did not terminate in 48 hours with optimizations disabled), we see a maximum speedup of 2.23X, compared to 17.8X reported in~\cite{DBLP:conf/sigsoft/KarimipourPCS23}.  Our local variables optimization on its own yields speedups of \locoptinfpeeduprange, sometimes larger than the core optimizations.  Fortunately, the techniques are complementary, together yielding the best speedups of \alloptinfpeeduprange.

\begin{tcolorbox}[width=\columnwidth, arc=3mm, boxsep=0.25mm]
  \textbf{RQ4:} \toolname has acceptable inference performance, aided significantly by our new optimization.
  \end{tcolorbox}

\subsection{Assessing annotations}\label{sec:assessing-annotations}
\begin{table*}[t]
  \scriptsize
  \centering
  \renewcommand{\arraystretch}{1.3}
  \resizebox{\textwidth}{!}{
    \begin{tabular}{l|rrrrrr|rrrr}
    \toprule
    \multirow{2}{*}{Project}  & \multicolumn{3}{c}{Manual Annotation Count} & \multicolumn{3}{c|}{Inferred Annotation Count} & \multicolumn{4}{c}{Error Count} \\ \cline{2-11}
                              & Total & TypeArg & PolyTaint & Total & TypeArg & PolyTaint & C off & G off & P off & All Active \\ \hline
        esapi-java-legacy     & 278   & 20      & 34        & 265   & 13      & 34        & 49    & 16    & 13    & 13         \\ \hline
        pybbs                 & 95    & 27      & 4         & 105   & 46      & 4         & 21    & 16    & 2     & 2          \\ \hline
        alfresco-core         & 81    & 38      & 0         & 81    & 36      & 0         & 12    & 5     & 2     & 2          \\ \hline
        alfresco-remote-api   & 110   & 12      & 10        & 213   & 22      & 7         & 66    & 42    & 27    & 24         \\ \hline
        cxf                   & 380   & 62      & 37        & 296   & 51      & 42        & 73    & 37    & 20    & 11         \\ \hline
        struts                & 347   & 50      & 5         & 313   & 47      & 7         & 116   & 54    & 44    & 37         \\ \hline
        commons-configuration & 239   & 10      & 25        & 261   & 12      & 23        & 33    & 17    & 11    & 11         \\
    \bottomrule
    \end{tabular}
  }
  \caption{Number of annotations inserted manually and by \toolname inference, and final error counts with various features disabled; C for checker improvements, G for generics inference, and P for \polytaint inference.}
  \label{tbl:humanvann-ablation}
\end{table*}

\Cref{table:oss:info} shows the number of annotations inferred by \toolname for our benchmarks and the corresponding annotation density, addressing RQ4.  The number of inferred annotations per KLoC is relatively low, ranging from 2.5-14.4.  In comparison, Banerjee et al.~\cite{banerjee19nullaway} reported an average of 13 annotations per KLoC for NullAway, ranging up to 46 annotations, and that checker has been widely adopted.

\begin{tcolorbox}[width=\columnwidth, arc=3mm, boxsep=0.25mm]
\textbf{RQ5:} The annotation requirements of \toolname are low enough to enable adoption, given the importance of preventing tainting vulnerabilities.
\end{tcolorbox}

\Cref{tbl:humanvann-ablation} compares the number of annotations inferred by \toolname to the number of annotations inserted in a separate manual process.  Two co-authors added manual annotations, each checking the other's work and coming to consensus for any disagreement.  We completed this process, adding 1,530 manual annotations.  We limited the manual changes to inserting annotations, prohibiting code changes.  This limitation is contrived: a developer would most likely fix bugs and refactor code alongside adding annotations.  We chose this methodology to fairly compare with \toolname inference, which only inserts annotations.

\Cref{tbl:humanvann-ablation} shows that the number of inserted annotations does tend to differ between the two approaches.  We inspected the discrepancies in detail, and found that in all cases, \toolname's inserted annotations were reasonable; which annotations were ``better'' was a subjective question.  In cases where \toolname inserted fewer annotations, one pattern was where a manually-written annotation captured some desired invariant, but \toolname eschewed the annotation since it increased the final error count.  A second pattern were cases where during manual annotation, an opportunity to use \polytaint was missed, but \toolname made use of \polytaint to avoid several other \untaintannot annotations.

For cases where \toolname inserted more annotations, a common explanation was again reducing error count; \toolname would insert many extra annotations to reduce the final error count by one, but this was not deemed to be worthwhile during manual annotation.
In such cases, arguably a better fix would be to restructure the code so fewer annotations would be needed.  We could easily add a setting to \toolname to limit the number of annotations it inserts to fix a single warning to avoid such cases.  In\ifextended{ Appendix~\ref{sec:manual-vs-inferred-examples}}\else{ an extended version of the paper~\cite{extendedPaperVersion}}\fi, we give detailed examples illustrating the above scenarios.

\begin{tcolorbox}[width=\columnwidth, arc=3mm, boxsep=0.25mm]
\textbf{RQ6:} \toolname's inserted annotations were always acceptable for insertion into source code, and sometimes improved on our manual annotations.
\end{tcolorbox}

\subsection{Ablation}\label{sec:ablation}

Finally, to answer RQ6, we performed an ablation study to see how many errors are reported by \toolname when disabling the checker features of \Cref{sec:checker-improvements}, generics inference (\Cref{sec:generic-fix-computation}), and \polytaint inference (\Cref{sec:polytaint-fixes}) individually.  The results are shown in \Cref{tbl:humanvann-ablation}.  We see large increases in reported errors with our new checker features disabled (3X-10.5X more errors), and similar impacts with generics support disabled (1.2X-8X more errors), showing the criticality of these features for precision.
Further, without the new checker features, 85 of the additional errors were false positives that could not be removed with annotations (see \Cref{sec:other-language-constructs}); such cases require a warning suppression, causing developer frustration.

Inference of \polytaint has a less significant impact on final error counts.  However, \polytaint inference is still critical for generating annotations close to what would be manually written, as shown in \Cref{tbl:humanvann-ablation}.  Manual annotations included a total of 115 \polytaint annotations across our benchmarks, and with inference of such annotations disabled, none of these would be found by \toolname.  Similarly, \Cref{tbl:humanvann-ablation} shows that generic type annotations are used commonly, and hence \toolname's support for inferring these annotations is important.

\begin{tcolorbox}[width=\columnwidth, arc=3mm, boxsep=0.25mm]
  \textbf{RQ7:} Our new checker features, generic type support, and \polytaint support were all critical to \toolname's effectiveness.
\end{tcolorbox}

\subsection{Threats to Validity}

Our benchmark choices are a threat to external validity.  We described our methodology for choosing benchmarks in \Cref{sec:experimental-setup}.  While we strove to choose diverse benchmarks, it is possible that \toolname will be less effective on a different set of programs.  Our choice of tools for comparison (CodeQL and P/Taint) is another threat to external validity; other whole-program static analyzers may perform differently.  See \Cref{sec:experimental-setup} for our process in choosing these tools.  Paper co-authors performing the manual annotation of benchmarks is a threat to internal validity.  We strove to add these annotations disregarding the workings of \toolname's inference.  Other developers may manually annotate code in different ways, but a user study evaluating the degree of such differences is out of scope for this work.  Implementation bugs in \toolname may also impact internal validity; but, we have done significant checking of correctness using manual inspection and a suite of regression tests.

\section{Related Work}

There is a broad literature on taint analysis; here we discuss the most closely related work.  In \Cref{sec:background} we contrasted type-based techniques with whole-program approaches~\cite{Banerjee2023compositional,tripp09taj,arzt14flowdroid,DBLP:conf/issta/HuangDMD15,DBLP:journals/pacmpl/GrechS17,codeqlmain,DBLP:conf/sigsoft/EmmiHJPRSSV21,DBLP:conf/sigsoft/WangWZYG020,DBLP:conf/uss/LivshitsL05,DBLP:conf/icse/WassermannS08,DBLP:conf/fase/HuangDM14}, and we compared with CodeQL~\cite{codeqlmain} and P/Taint~\cite{DBLP:journals/pacmpl/GrechS17} in our evaluation.  Recent work by Banerjee et al.~\cite{Banerjee2023compositional}  describes an approach to incremental  taint analysis but does not evaluate its performance.
Szabo~\cite{DBLP:conf/sigsoft/Szabo23} presents an initial exploration of incrementalizing CodeQL analyses.  While their incremental running times were promising, the additional memory usage of their technique was prohibitive.  Type-based taint analysis is naturally incremental and does not suffer from these engineering challenges.

The most closely related whole-program approach is that of Huang et al.~\cite{DBLP:conf/issta/HuangDMD15}, which partially inspired our work.  Their work is also formulated in terms of type-based taint analysis and type inference.  Their type system does not support generic types: instead, they apply polymorphic types to fields to achieve a form of field sensitivity.  It is unclear how to expose such field polymorphism in terms of standard pluggable types.  Their work does not present a technique to persist types into source code to enable checking without inference and is specific to Android applications so we did not include this tool in our evaluations.
\toolname performs inference for standard Java pluggable types, including generic types, enabling a straightforward integration with standard development workflows.
The inference of~\cite{DBLP:conf/issta/HuangDMD15} runs faster than ours since it operates over a single constraint encoding.  However, our inference works with an existing checker without requiring a constraint encoding, which simplifies making improvements to the checker like those of \Cref{sec:checker-improvements}.

There are many other approaches to enforcing information flow properties with types.  The well-known Jif system~\cite{10.1145/292540.292561} and many subsequent works support sophisticated properties like multiple principals.  Such features are not necessary to capture the vulnerabilities targeted by typical taint analyses and this work.  Ernst et al.~\cite{DBLP:conf/ccs/ErnstJMDPRKBBHVW14} target verification of Android apps, also using pluggable types.  Their system aims for soundness for a security-critical military context, leading to an annotation burden of ~60 annotations per KLoC, much higher than ours.  \toolname eschews strict soundness to reduce the annotation burden for usability.  We believe \toolname's inference technique could be adapted to systems like~\cite{DBLP:conf/ccs/ErnstJMDPRKBBHVW14} in the future.

Regarding alternate inference approaches, standard type inference~\cite[Chapter 22]{10.5555/509043} tries to discover a complete typing for a program in a given type system.  Such techniques do not directly apply to our scenario, as a typical program will not be verifiable in our taint type system solely through adding annotations, due to true or false positives; we aim to find a good set of annotations for such untypable programs.  Kellogg et al.~\cite{DBLP:conf/kbse/KelloggDNAE23} present a general inference technique for any pluggable type system built on the Checker Framework.  However, their technique does not infer polymorphic or generic type annotations, and it may generate many more annotations than what would be written by developer.

Checker Framework Inference~\cite{cf_inference} uses constraint-based approach to infer types, with applications to domains like a type system for measurement units~\cite{DBLP:journals/pacmpl/XiangLD20}.  We chose a ``black box'' inference approach~\cite{DBLP:conf/sigsoft/KarimipourPCS23} for this work since it enabled re-using a checker implementation without reimplementing its logic in a constraint language.  The approach of~\cite{DBLP:journals/pacmpl/XiangLD20} does not support inference of polymorphic method annotations like \polytaint, and the paper does not discuss in detail its level of support for inferring annotations on generic type arguments.

\section{Conclusions}

We have presented \toolname, a novel approach to type-based taint checking and inference for Java.  \toolname includes a novel checker that makes the core tainting type system more practical to use, and a novel inference algorithm capable of handling generic types and polymorphic annotations.  Our evaluation showed that \toolname provided significant scalability advantages over a standard approach, with improved recall and comparable precision, and inferred annotations suitable for direct inclusion in source code.  Hence, \toolname is a significant step toward practical and widescale type-based taint checking for Java.

\clearpage

\bibliographystyle{plainurl}
\bibliography{refs}

\ifextended
\clearpage

\markboth{Appendix}{Appendix}
\appendix

\section{Manual vs. Inferred Annotations Examples}\label{sec:manual-vs-inferred-examples}

Here we give some illustrative examples showing how our manual annotations in our experimental evaluation sometimes differed from those inserted by \toolname's inference.

\begin{figure}[t]
  \centering
  \begin{subfigure}[b]{1.0\textwidth}
    \begin{lstlisting}
Map<String, $\color{green!50!black}\texttt{+@Untainted}$ String> map;
void add(String key, $\color{green!50!black}\texttt{+@Untainted}$ Stringval) {
  map.put(key, val);
}
$\color{green!50!black}\texttt{+@Untainted}$ String getValue(String key) {
  return map.get(key);
}
void foo(String key) {
  sink(getValue(key)); $\color{red!50!black}// \; initial \; error$
}
void bar(TaintSource source) {
  add("key1", source.getTaintedData()); $\color{red!50!black}// \; remaining \; error$
}
void baz(TaintSource source) {
  add("key2", source.getTaintedData()); $\color{red!50!black}// \; remaining \; error$
}\end{lstlisting}
    \caption{Human annotated code}
    \label{fig:annotations_insert_cmp_human_human}
  \end{subfigure}
  \hfill
  \begin{subfigure}[b]{1.0\textwidth}
    \begin{lstlisting}
Map<String, String> map;
void add(String key, String val) {
  map.put(key, val);
}
$\color{green!50!black}\texttt{+@Untainted}$ String getValue(String key) {
  return map.get(key); $\color{red!50!black}// \; remaining \; error$
}
void foo(String key) {
  sink(getValue(key)); $\color{red!50!black}// \; initial \; error$
}
void bar(TaintSource source) {
  add("key1", source.getTaintedData());
}
void baz(TaintSource source) {
  add("key2", source.getTaintedData());
}\end{lstlisting}
    \caption{Inference Annotated Code}
    \label{fig:annotations_insert_cmp_human_annotator}
  \end{subfigure}
  \caption{Example for how inference annotated code and human annotated code compares. Green text indicates annotations inserted by inference and human annotator.}
  \label{fig:annotations_insert_cmp_human}
\end{figure}

In \Cref{fig:annotations_insert_cmp_human_human}, the code presents a real taint issue where a tainted data-flow is possible. At line 9 of Figure \ref*{fig:annotations_insert_cmp_human_human}, \texttt{getValue(key)} is passed to a sink that expects an untainted data. Disregarding any inserted green annotations, our checker will find it to be tainted and report an error. While inserting 
annotations to remove this error, the human annotator traces back starting from the sink. They will find that making the return type of method \texttt{getValue} untainted will remove the error. But the value is retrieved from field \texttt{map}, so relevant type argument needs to be untainted as well. Method \texttt{getValue} gets \texttt{value} from the second type argument to \texttt{map}. Now at line 11 and line 12, taint sources are adding values to the \texttt{map} field. These can not be resolved by adding any more annotatations. So this code will have 2 remaining errors after inserting 3 annotations by the human annotator.

In \Cref{fig:annotations_insert_cmp_human_annotator}, \toolname decides that adding only one annotation at line 5 ends up with 1 error, and adding further annotations increases the number of errors. Thus it decides to stop at this point. This is a good example of how the \toolname may insert fewer annotations than the manual process in its goal of minimizing the error count.

\begin{figure}[t]
  \centering
  \begin{subfigure}[b]{1.0\textwidth}
    \begin{lstlisting}
$\color{green!50!black}\texttt{+@Untainted}$ String get($\color{green!50!black}\texttt{+@Untainted}$ Properties prop, 
$\color{green!50!black}\texttt{+@Untainted}$ String key) {
  return prop.getProperty(key);
}
void foo($\color{green!50!black}\texttt{+@Untainted}$ Properties prop, 
$\color{green!50!black}\texttt{+@Untainted}$ String key) {
  sink(get(prop, key)); $\color{red!50!black}// \; initial \; error$
}
void bar($\color{green!50!black}\texttt{+@Untainted}$ Properties prop, 
$\color{green!50!black}\texttt{+@Untainted}$ String key) {
  get(prop, key);
}
void baz($\color{green!50!black}\texttt{+@Untainted}$ Properties prop, 
$\color{green!50!black}\texttt{+@Untainted}$ String key) {
  get(prop, key);
}\end{lstlisting}
    \caption{Human annotated code}
    \label{fig:annotations_insert_cmp_human_human_polytaint}
  \end{subfigure}
  \hfill
  \begin{subfigure}[b]{1.0\textwidth}
    \begin{lstlisting}
$\color{green!50!black}\texttt{+@PolyTaint}$ String get($\color{green!50!black}\texttt{+@PolyTaint}$ Properties prop, 
$\color{green!50!black}\texttt{+@PolyTaint}$ String key) {
  return prop.getProperty(key);
}
void foo($\color{green!50!black}\texttt{+@Untainted}$ Properties prop, 
$\color{green!50!black}\texttt{+@Untainted}$ String key) {
  sink(get(prop, key)); $\color{red!50!black}// \; initial \; error$
}
void bar(Properties prop, String key) {
  get(prop, key);
}
void baz(Properties prop, String key) {
  get(prop, key);
}\end{lstlisting}
    \caption{Inference Annotated Code}
    \label{fig:annotations_insert_cmp_human_annotator_polytaint}
  \end{subfigure}
  \caption{Example for how inference annotated code and human annotated code compares with polytaint annotatations. Green text indicates annotations inserted by inference and human annotator.}
  \label{fig:annotations_insert_cmp_human_polytaint}
\end{figure}

Another example of a difference is presented in \Cref{fig:annotations_insert_cmp_human_polytaint}. In \Cref{fig:annotations_insert_cmp_human_human_polytaint}, the checker initially reports an error at line 7, where method \texttt{get} returns tainted value to \texttt{sink}. The human annotator finds that \texttt{getProperty} is an invocation to unannotated code, so they make both the receiver and argument at line 3 \untaintannot. Now they also need to make arguments to method \texttt{foo} untainted. But this has further implications. Even though \texttt{get} is called from method \texttt{bar} and method \texttt{baz}, the return value is not passed to any sink. However, since \texttt{get} expects untainted arguments, the human annotator needs to make the arguments of these methods untainted. If there are more methods calling \texttt{get}, the annotator will have to untaint the arguments passed to \texttt{get} from those methods as well. This can result in a large number of annotations.
In contrast, the inference algorithm infers \polytaint annotations and untaints the arguments of method \texttt{foo}. It does not need to untaint the arguments of methods \texttt{bar} and \texttt{baz} as they do not pass the return value to any sink. 
While no errors are reported at the ends for both the human annotated code and the inference annotated code, the number of annotations inserted by the human annotator can be significantly higher than the number of annotations inserted by the inference algorithm, due to the missed opportunity to use \polytaint.

\begin{figure}[H]
  \centering
  \begin{subfigure}[b]{1.0\textwidth}
    \begin{lstlisting}
foo(TaintSource source) {
  List<String> tainted = source.getTaintData();
  sink1(tainted);
  sink2(tainted);
}\end{lstlisting}
    \caption{Human annotated code}
    \label{fig:annotations_insert_cmp_human_additional_human}
  \end{subfigure}
  \hfill
  \begin{subfigure}[b]{1.0\textwidth}
    \begin{lstlisting}
foo(TaintSource source) {
  List<$\color{green!50!black}\texttt{+@Untainted}$ String> tainted = source.getTaintData(); 
  sink1(tainted);
  sink2(tainted);
}\end{lstlisting}
    \caption{Inference Annotated Code}
    \label{fig:annotations_insert_cmp_human_additional_annotator}
  \end{subfigure}
  \caption{Example for how inference annotated code and human annotated code compares. Green text indicates annotations inserted by inference and human annotator.}
  \label{fig:annotations_insert_cmp_human_additional}
\end{figure}

As shown in the code example in Figure \ref*{fig:annotations_insert_cmp_human_additional}, it can also be the case that annotator infers more annotations than a human annotator. In Figure \ref*{fig:annotations_insert_cmp_human_additional_human}, \toolname initially reports two errors where \texttt{tainted} is passed to two sinks at line 3 and line 4. A human annotator may decide to not add any annotatations since the taint flow is obvious and presents a real issue. However, \toolname decides to add \untaintannot annotation to the type argument of \texttt{List} at line 2 in Figure \ref*{fig:annotations_insert_cmp_human_additional_annotator}. This removes both errors but consolidates them into one single error at the assignment at line 2. This is an example of how \toolname can end up with more annotations than a human annotator. Such scenarios are more prominent in benchmark pybbs, alfresco-remote-api, and commons-configuration in Table \ref*{tbl:humanvann-ablation}, where the annotator infers more annotations than the human annotator. 
\fi

\end{document}